\documentclass[aps,prb,reprint,groupedaddress,showpacs,nofootinbib]{revtex4-1}
\usepackage{caption}
\usepackage{bm}
\usepackage{amsmath}
\usepackage[cp1251]{inputenc}

\input{epsf}

\begin{document}
\title{Electron states on the smooth edge of 2D topological insulator: elastic backscattering and light absorption}

\author{M.M. Mahmoodian}
\email{mahmood@isp.nsc.ru}
\affiliation{Rzhanov Institute of Semiconductor Physics, Siberian Branch, Russian Academy of Sciences, Novosibirsk, 630090 Russia}
\affiliation{Novosibirsk State University, Novosibirsk, 630090 Russia}

\author{M.V. Entin}
\email{entin@isp.nsc.ru}
\affiliation{Rzhanov Institute of Semiconductor Physics, Siberian Branch, Russian Academy of Sciences, Novosibirsk, 630090 Russia}
\affiliation{Novosibirsk State University, Novosibirsk, 630090 Russia}

\date{\today}

\begin{abstract}

The 2D TI edge states are considered within the Volkov-Pankratov (VP) Hamiltonian. A smooth transition between TI and OI is assumed. The edge states are formed in the total gap of homogeneous 2D material. A pair of these states are of linear dispersion, others have gapped Dirac spectra. The optical selection rules are found. The optical transitions between the neighboring edge states appear in the global 2D gap for the in-plane light electric field directed across the edge.

The electrons in linear edge states have no backscattering, that is indicative of the fact of topological protection. However, when linear edge states get to the energy domain of Dirac edge states, the backscattering becomes permitted. The elastic backscattering rate is found. The Drude-like conductivity is found when the Fermi level gets into the energy domain of the coexistence of linear and Dirac edge states. The localization edge conductance of a finite sample at zero temperature is determined.

\end{abstract}

%\pacs{73.21.Fg, 73.61.Ga, 73.22.Gk, 72.25.Dc}

\maketitle
\section{Introduction}

The systems under consideration are 2D topological insulators (TI) \cite{hasan,fu2,konig,qi} (more references can be found in \cite{hasan,qi}). The most intriguing property of them is the internal insulating part with conducting edges. These systems were been first invented by Bernevig-Hughes-Zhang (BHZ)\cite{bhz}. However, earlier, a similar 3D TI had been studied by Volkov-Pankratov (VP) \cite{vp}. The edge states of 2D topological insulators attract attention due to their possibility of the topological protection of electrons from backscattering processes. The difference between VP and BHZ models is that BHZ is applied to the external edge of the TI, while VP can be applicable to the interface between a TI and an ordinary insulator. Many unusual features of the 2D TI edge states were studied. In particular, the non-local conductance due to topologically-protected linear-edge states was  observed in \cite{molenkamp,kvon}, and theoretically discussed by  \cite{buhmann} and; however, the non-locality does not obligatorily manifest itself. The edge state light absorption was theoretically \cite{ent-mag,mah-mag-ent,mah-ent} and experimentally investigated \cite{kvon1}. Different theoretical approaches to this intriguing problem were developed. In particular, we found that the topologically protected edge states possess a unique property: electrons in these states do not interact with the external electric field and can not scatter by each other \cite{ent-brag1}.

If the TI boundary is sharp, the only topologically protected edge states exist. However, the TI boundary is not necessarily sharp. In this case the additional edge states are formed \cite{vp}, \cite{tchou,lu}. The gap sign, which determines if the 2D layer is OI or TI, is controlled by the layer thickness \cite{mah-ent,mah-ent1}. Due to intentional or non-intentional a smooth transition between TI and OI domains are formed near formal TI boundary.

The smooth transition between an OI with a positive and a TI with a negative gap generates not only the edge states with linear energy spectrum $\pm vk_x$ (here and in what follows, $\hbar=1$), where $v$ and $k$ are the velocity and the momentum of electron on the edge state, but the series of the Dirac-like states $\pm\sqrt{v^2k_x^2+\epsilon_n^2}$. These states cover the 2D energy gap. Unlike linear edge states, Dirac-like edge states have no topological protection. Under  illumination they can provide light absorption. The purpose of the present paper is to find this absorption and the elastic backscattering of electrons with linear spectra with the participation of Dirac edge states.

Both planned problems need the knowledge of the matrix elements between the edge states due to the interaction with light for absorption, or due to the interaction with impurities for elastic backscattering. Hence, these problems can be studied in common.

First, we will formulate the problem. Then, the wave functions in the model with gap $\propto \tanh(z/a)$ will be presented. After that, we will present the case of linear transition between the domains with a negative and positive gap. This is a limiting case of the previous model, which is applicable also to more general models of smooth transition. Then the elastic backscattering rate will be found. After that the optical absorption, due to transitions between linear and the Dirac states, will be found. The results will be summarized in the Discussion section. The calculation details are in the Appendix.

\subsection*{Volkov-Pankratov Hamiltonian in 2D}

We will base on the VP Hamiltonian for 2D TI \cite{vp}. The 2D system is located in plane $(x,z)$. The transition between a TI and an OI is modeled by the one-dimensional dependence of gap $\Delta(z)=\Delta_0F(z/l)$, where $\Delta_0$ is half of the bulk gap so that dimensionless function $F(z>0)>0$ and $F(z<0)<0$. In particular, one can assume that $F(z)=\tanh(z)$. If parameter $l\to 0$, the TI-OI transition is step-like. If $l$ is large, the transition is smooth.

The two-dimensional VP Hamiltonian reads \cite{vp}
\begin{eqnarray}\label{g0}
H_0=-\Delta(z)\tau_y + v\tau_x\left(k_x\sigma_x+k_z\sigma_z\right),
\end{eqnarray}
where $\bm{\tau}$ and $\bm{\sigma}$ are Pauli matrices, which act on orbital and spin subspaces, respectively.

The energy spectrum of Hamiltonian (\ref{f1}) for the TI-OI transition $\Delta(z)=\Delta_0\tanh(z/l)$ \cite{vp}
\begin{eqnarray}\label{g1}
&&E_{n,\sigma}^{\lambda}(k_x)=\lambda\sqrt{v^2k_x^2+\epsilon_{n,\sigma}^2},\\
&&\epsilon_{n,\sigma}^2=\Delta_0^2\left[ 1 - \left( 1- \left( n + \frac{1+\sigma}{2} \right) \frac{v}{\Delta_0 l} \right)^2 \right],
\end{eqnarray}
where $n$ are integers, $\lambda=\pm$, and $0\leq n + (1+\sigma)/2\leq \Delta_0 l/v$.

\begin{figure}[ht]
\centerline{\epsfysize=6cm\epsfbox{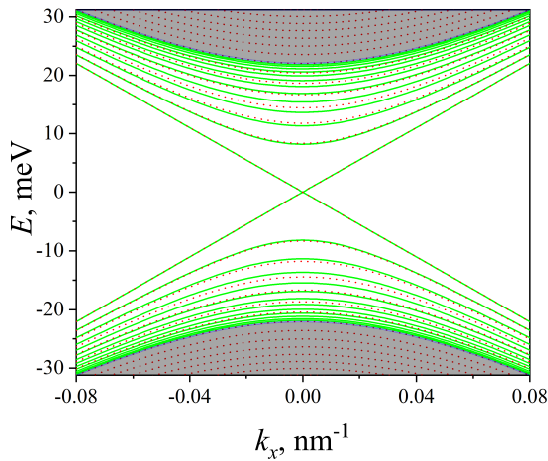}}
\caption{Edge states for $\Delta(z)=\Delta_0\tanh(z/l)$ (green, solid) and for $\Delta(z)=\Delta_0 (z/l)$ (red, dashed). The chosen parameters are $\Delta_0=22$ meV and $v=4.2*10^7\mbox{cm/s}$, and that corresponds to HgTe layer widths $55$~{\AA} and $67.6$~{\AA}, and $l=1750.5$~{\AA}. The lowest exact and approximate states coincide. At large numbers the exact states condense when approaching the boundary of 2D states (filled, light blue).}\label{fig1}
\end{figure}

For $n=0, \sigma=-1$ we have the Weyl branch with linear spectrum $E_{0,-1}^{\lambda}=\lambda vk_x$. The other Dirac-like branches with $n,\sigma=1$ and $n+1,\sigma=-1$, have gaps. These states are double-degenerate (see Fig.~\ref{fig1}).

\subsection*{Elastic backscattering of edge-state-electrons in a two-dimensional topological insulator}
The presence of Dirac edge states inside the characteristic gap of two-dimensional edge states is due to the smooth transition between the positive and negative gaps. A typical transition scale is described by letter $l$.

The two-dimensional gap $2\Delta_0$. The typical distance to the first Dirac state is $\sqrt{\Delta_0v/l}$, where $v$ is the Fermi velocity; the linear branches are of the form $\pm vk_x$, which is less than $\Delta_0$ for $\Delta_0\gg v/l$. With an abrupt transition, only a linear branch remains inside the two-dimensional gap. When $l$ is very large, the transition can be replaced by a linear dependence $\Delta=\Delta_0z/l$. This dependence leads to an exactly solvable problem. The spectrum consists of a pair of linear branches $\pm vk_x$ and Dirac branches.

In the narrow energy range $|E|<\sqrt{\Delta_0v/l}$, there are only linear topologically protected branches. However, outside the narrow region, these states overlap in energy with Dirac states. At low temperatures, the elastic transition processes between linear and Dirac and between Dirac states are allowed. This gives the backscattering of electrons, including those on linear branches, that is, from a state with a velocity $v$ with momentum $k_x$ to a state with a velocity $–v$ with momentum $-k_x$. The expected process is two-step: $vk_x\to\sqrt{v^2k_x^2+\epsilon_{n,\sigma}^2}\to(-v)(-k_x)$.

Consider the probability of transition between states $n,k$ and $n',k'$ under the action of $\exp (iq_zz + iq_xx)$. Matrix element $J=\langle n'k_x'| \exp(iq_zz+ iq_xx)|n,k_x\rangle$ is proportional to $\delta(k_x'-k_x-q_x)$. The square $|J|^2$ must be multiplied by the square of the Fourier transform of the impurity potential $|V({\bf q})|^2$ and the impurity concentration $n_i$. For Coulomb impurities
\begin{eqnarray}\label{g2}
V({\bf q})=\frac{2\pi e^2}{\kappa q}\left(1-e^{-2qd}\right),
\end{eqnarray}
where $d$ is the distance to the metallic gate. As $q\to 0$, the Fourier transform of the potential tends to a constant.

In a general case, the inverse transition time is
\begin{eqnarray}\label{g3}
\frac{1}{\tau}=n_i\int\limits_{-\infty}^\infty dq_z|V({\bf q})|^2|J|^2|_{k_x'-k_x=q_x}g_{n'}(E_{n',\sigma,k_x'}^{\lambda})dE_{n',\sigma',k_x'}^{\lambda}.
\end{eqnarray}

The density of states has the form
\begin{eqnarray}\label{g4}
g_n(E)=\frac{E}{\pi v\sqrt{E^2-\epsilon_{n,\sigma}^2}}.
\end{eqnarray}

The presence of the transitions indicated here leads to a limitation of the topological security of a linear spectrum.

\begin{figure}[ht]
\centerline{\epsfysize=6cm\epsfbox{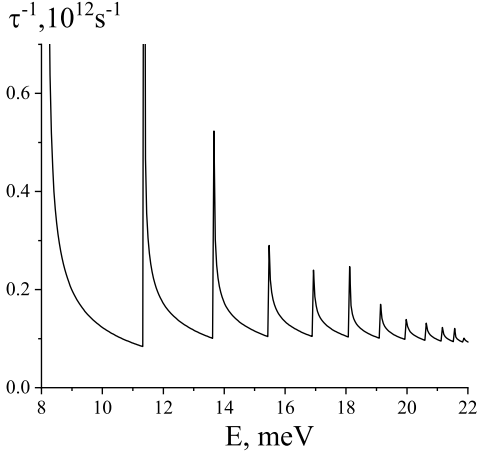}}
\caption{Energy dependence of the inverse transition time for transitions $E_{0,-1}^+=vk_x$ $\rightarrow$ $E_{n',\sigma}^+(k_x)$. The chosen parameters are $\Delta_0=22$ meV, $v=4.2*10^7\mbox{cm/s}$, $l=1750.5$~{\AA}, $d=1000$~{\AA}, $\kappa=10.3$, $n_i=10^9\mbox{cm}^{-2}$, respectively}\label{fig2}
\end{figure}

In Fig.~\ref{fig2} is the energy dependence of the inverse transition time in the region where the linear edge states overlap the first Dirac subband.

\section{Light absorption in a two-dimensional topological insulator}

Later on we shall deal with a smooth transition. When the edge state size $\xi=v/\Delta_0$ is less than the transition width $l$, the dependence $\Delta(z)$ may be expanded as $\Delta(z)=\Delta_0z/l$. In such case the edge states can be expressed via oscillator wave functions.

Here it is convenient to use a variant of the Hamiltonian that was originally proposed by Zhang et al. \cite{fzhang}
\begin{eqnarray}\label{f1}
H_0=\Delta(z)\tau_z + v\left(k_z\tau_y-k_x\tau_x\sigma_y\right).
\end{eqnarray}

After unitary transformation $T=\exp(i\pi\tau_y/4)$, the Hamiltonian (\ref{f1}) with $\Delta(z)=\Delta_0z/l$ can be rewritten as \cite{tchou,lu}
\begin{eqnarray}\label{f2}
H_T=TH_0T^{\dag}=v\left[
               \begin{array}{cccc}
                   0 & ik_x & \frac{\sqrt{2}}{l_0}\hat{c} & 0 \\
                   -ik_x & 0 & 0 & \frac{\sqrt{2}}{l_0}\hat{c} \\
                   \frac{\sqrt{2}}{l_0}\hat{c}^{\dag} & 0 & 0 & -ik_x \\
                   0 & \frac{\sqrt{2}}{l_0}\hat{c}^{\dag} & ik_x & 0 \\
               \end{array}
         \right],
\end{eqnarray}
where the ladder operators are
\begin{eqnarray}\label{f3}
\hat{c}=\frac{l_0}{\sqrt{2}}\left(\frac{z}{l_0^2}+ik_z\right) ~~~\mbox{and}~~~ \hat{c}^{\dag}=\frac{l_0}{\sqrt{2}}\left(\frac{z}{l_0^2}-ik_z\right),
\end{eqnarray}
which satisfy the usual commutation relations $[\hat{c},\hat{c}^{\dag}]=1$, $l_0=\sqrt{l\xi}$. The Hamiltonian (\ref{f2}) has the energy spectrum
\begin{eqnarray}\label{f4}
E_n^{\lambda}(k_x)=\lambda v\sqrt{k_x^2+\frac{2n}{l_0^2}},
\end{eqnarray}
where $\lambda=\pm$. The case $n=0$ corresponds to the edge states with linear dispersion, and $n>0$ represents massive VP states.

The Kubo formula for light polarized along the $z$-direction is
\begin{eqnarray}\label{f5}
\sigma_{zz}(\omega)=i e^2\sum\limits_{\substack{m,n\in N\\ \lambda,\lambda'}} \int\limits_{-\infty}^{\infty}\frac{dk_x}{2\pi}
\frac{|\langle\psi_m^{\lambda'}|\hat{v}_z|\psi_n^{\lambda}\rangle|^2}{E_m^{\lambda'}-E_n^{\lambda}}\times\nonumber\\
\frac{f(E_n^{\lambda})-f(E_m^{\lambda'})}{E_m^{\lambda'}-E_n^{\lambda}-\omega+i\delta}.
\end{eqnarray}
Here, $f(E)$ is the Fermi-Dirac distribution function, $\hat{v}_z=v\tau_y$, $\delta\to+0$. For the light polarized along the $z$-direction, according to Eq.~(\ref{f11}), only the transitions $n\to n\pm 1$ are allowed. For $E_F>0$, three transition types can be distinguished. The transitions between the states with $\lambda=-$ and $\lambda=+$ are possible at high frequencies, starting from $\omega>\sqrt{2}v/l_0$. At frequencies below this value, transitions are possible only between the states with $\lambda=+$.
\begin{eqnarray}\label{f6}
\Re[\sigma_{zz}(\omega)]=\frac{e^2}{2\omega}\sum\limits_{\substack{m,n\in N\\ \lambda,\lambda'}} \int\limits_{-\infty}^{\infty}dk_x|\langle\psi_m^{\lambda'}|\hat{v}_z|\psi_n^{\lambda}\rangle|^2\times\nonumber\\
\delta\left(\omega-(E_m^{\lambda'}-E_n^{\lambda})\right)\left(f(E_n^{\lambda})-f(E_m^{\lambda'})\right).
\end{eqnarray}

\begin{figure}[ht]
\centerline{\epsfysize=5cm\epsfbox{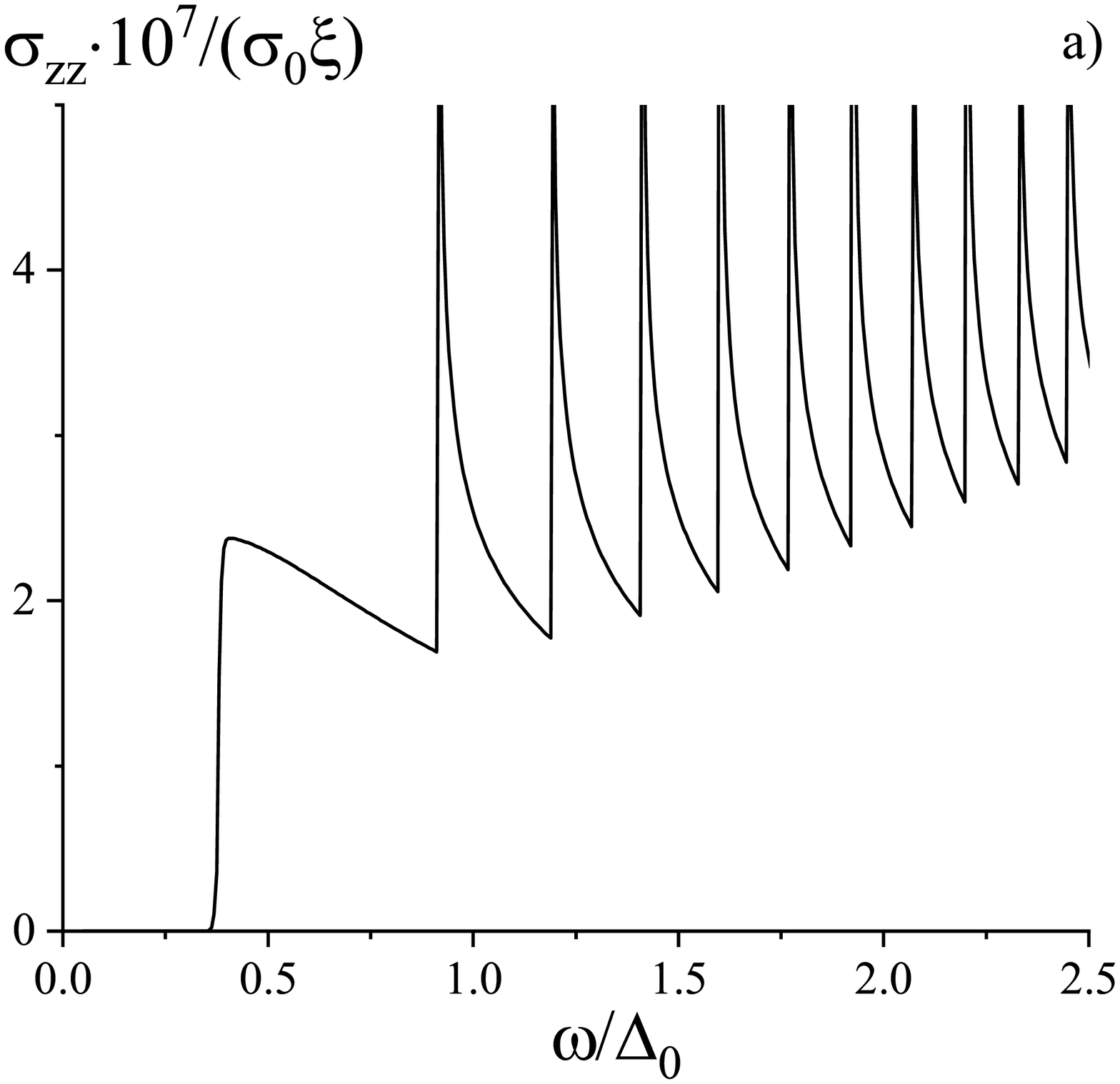}}
\centerline{\epsfysize=5cm\epsfbox{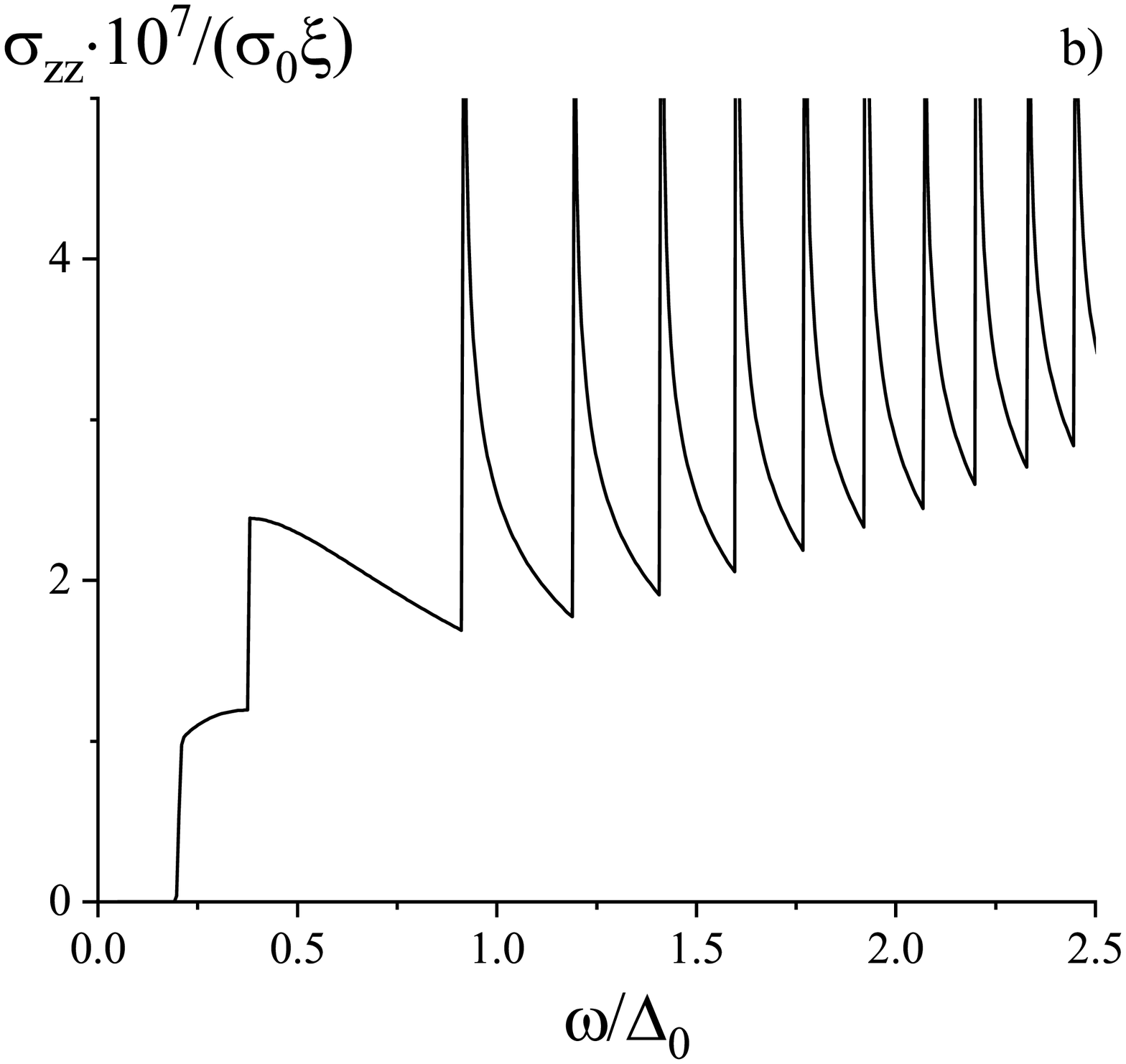}}
\centerline{\epsfysize=5cm\epsfbox{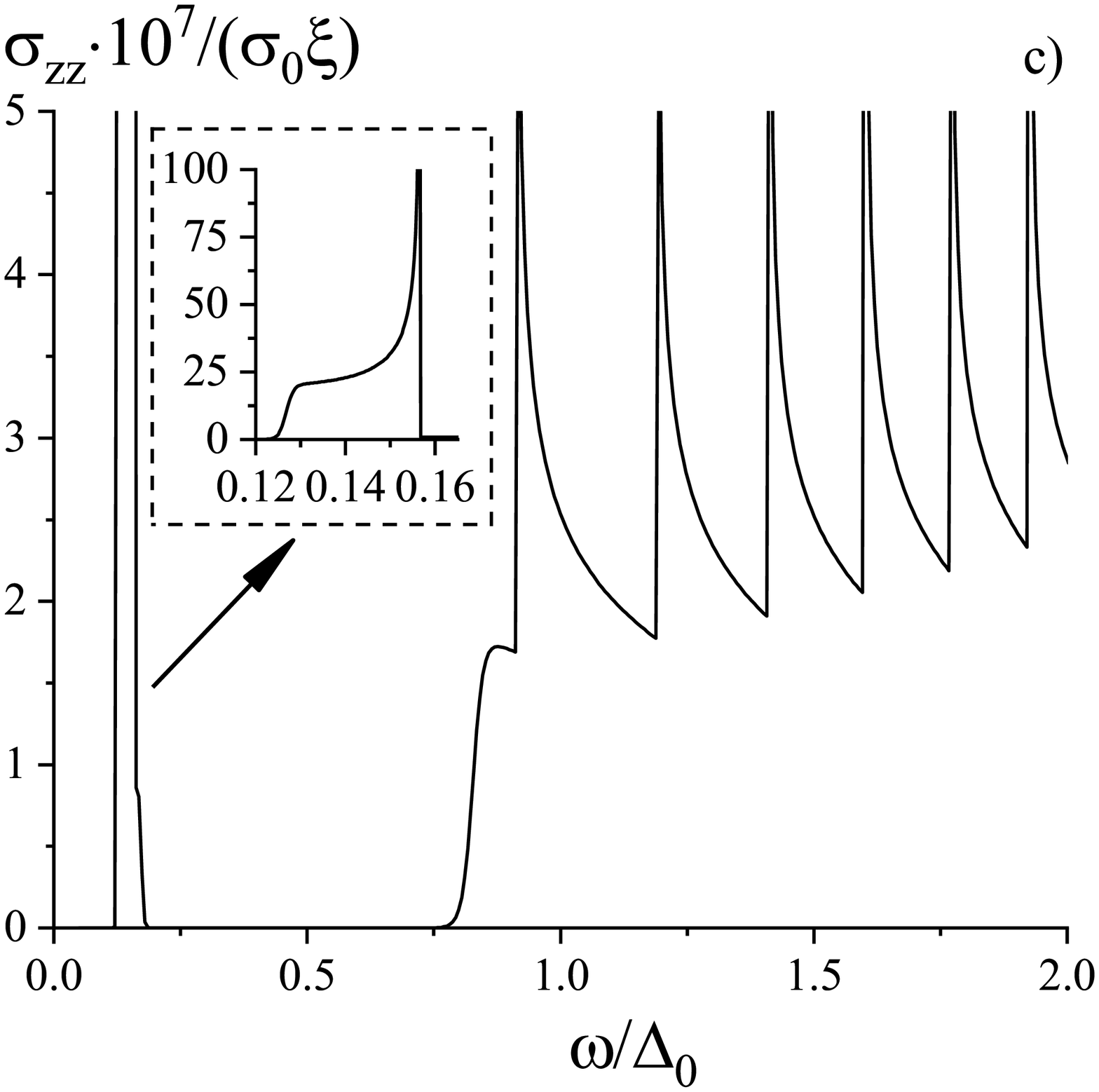}}
\caption{Optical conductivity $\Re[\sigma_{zz}]$ (in units of $\sigma_0\xi$, $\sigma_0=e^2/h$), as a function of the excitation energy $\hbar\omega$ (in units $\Delta_0$) for $E_F=0$ (Fig.~\ref{fig3}a), $E_F=0.25\Delta_0$ (Fig.~\ref{fig3}b) and $E_F=0.5\Delta_0$ (Fig.~\ref{fig3}c).}\label{fig3}
\end{figure}

In the Fig.~\ref{fig3} is the optical conductivity dependence $\Re[\sigma_{zz}(\omega)]$ on $\omega$ for different values of $E_F$.

\section{Conclusions}
We see that, in the absence of scattering, the selection rules allow transitions between the Dirac branches or between the Dirac and linear branches polarization of external in-plane microwave electric field across the edge. Besides, they do not allow the transitions between the linear branches for any external microwave electric field polarization. The transitions are allowed between neighboring transversal states. This statement is valid, however, in the assumption of linear dependence of the gap on the transversal coordinate, and that corresponds to the frequency much less than the gap value apart from edge $2\Delta_0$.

The selection rules determine the lower threshold for light absorption $\sqrt{\Delta_0/vl}$. This differs the case of smooth edge from the steep one, where the absorption is limited from below by the quantity $\Delta_0$ due to the processes of transitions between the edge and two-dimensional states.

The light absorption oscillates with the light frequency and the Fermi level, and has the $1/\sqrt{\omega-\omega_n}$ singularities at the thresholds of the pair density of states.

\paragraph*{\bf Acknowledgments.} This research was supported in part by the RFBR, grant No. 20-02-00622.

\section{Appendix I. Electrons states for $\Delta(z)=\Delta_0\tanh(z/l)$}

The eigenfunctions of the Hamiltonian (\ref{g0}) for $\Delta(z)=\Delta_0\tanh(z/l)$ are
\begin{eqnarray}\label{g_wf}
\psi_{n,1}^{\lambda}=\left( \frac{vk_x}{E_{n,\sigma}^{\lambda}}, ~\frac{i}{E_{n,\sigma}^{\lambda}} \left(\Delta+v\frac{\partial}{\partial z}\right), ~0, ~1 \right)\Psi_{n,\sigma}, ~~~n\geq 1,\nonumber\\
\psi_{n,2}^{\lambda}=\left( \frac{i}{E_{n,\sigma}^{\lambda}} \left(\Delta-v\frac{\partial}{\partial z}\right), \frac{vk_x}{E_{n,\sigma}^{\lambda}}, ~1, ~0 \right)\Psi_{n,\sigma}, ~~~n\geq 1,\nonumber\\
\psi_0^{\lambda}=\left( \lambda, ~ \frac{i\lambda}{vk_x} \left(\Delta+v\frac{\partial}{\partial z}\right), ~0, ~1 \right)\Psi_{n,\sigma}. ~~~n=0,\nonumber
\end{eqnarray}
Here $$\Psi_{n,\sigma}(\eta)=\frac{A_{n,\sigma}n!}{(\varepsilon+1)_n}\left(1-\eta^2\right)^{\varepsilon/2}P_n^{(\varepsilon,\varepsilon)}(\eta),$$
where $P_n^{(\varepsilon,\varepsilon)}(\eta)$ are Jacobi polynomials, $(\varepsilon+1)_n$ is the Pochhammer symbol, $\eta=\tanh(z/l)$ and $\varepsilon=\Delta_0 l/v-n-(1+\sigma)/2$, $A_{n,\sigma}$ are normalization constants.

\section{Appendix II. Electrons states and matrix elements for $\Delta(z)=\Delta_0z/l$}

The eigenstates of Hamiltonian (\ref{f2}) for $\Delta(z)=\Delta_0z/l$ are
\begin{eqnarray}\label{f7}
|\psi_n^{\lambda}\rangle=\left( a_{1,n}^{\lambda}|n-1\rangle, ~a_{2,n}^{\lambda}|n-1\rangle, ~a_{3,n}^{\lambda}|n\rangle, ~a_{4,n}^{\lambda}|n\rangle \right), ~~~n\geq 1,\nonumber\\
|\psi_0^{\lambda}\rangle=\left( 0, ~0, ~a_{3,0}^{\lambda}|0\rangle, ~a_{4,0}^{\lambda}|0\rangle \right), ~~~n=0,\nonumber
\end{eqnarray}
where $|n\rangle$ is the eigenstate of harmonic oscillator determined by $\hat{c}^{\dag}$ and $\hat{c}$. The Hamiltonian written in this basis reads:
\begin{eqnarray}\label{f8}
H_T(n)=v\left[
               \begin{array}{cccc}
                   0 & ik_x & \frac{\sqrt{2n}}{l_0} & 0 \\
                   -ik_x & 0 & 0 & \frac{\sqrt{2n}}{l_0} \\
                   \frac{\sqrt{2n}}{l_0} & 0 & 0 & -ik_x \\
                   0 & \frac{\sqrt{2n}}{l_0} & ik_x & 0 \\
               \end{array}
         \right],
\end{eqnarray}
The corresponding normalized eigenvectors of the Hamiltonian above are:
\begin{eqnarray}\label{f9}
\psi_{n,1}^{\lambda}=\frac{1}{\sqrt{2}}\left( i\cos\alpha_n, ~\lambda, ~0, ~\sin\alpha_n \right), ~~~n\geq 1,\\
\psi_{n,2}^{\lambda}=\frac{1}{\sqrt{2}}\left( \lambda, ~-i\cos\alpha_n, ~\sin\alpha_n, ~0, \right), ~~~n\geq 1,\\
\psi_0^{\lambda}=\frac{1}{\sqrt{2}}\left( 0, 0, i, -\lambda \right), ~~~n=0.
\end{eqnarray}
For $n\geq 1$,
\begin{eqnarray}\label{f10}
\cos\alpha_n=\frac{k_x}{\sqrt{k_x^2+\frac{2n}{l_0^2}}}, ~~~\sin\alpha_n=\frac{\frac{2n}{l_0^2}}{\sqrt{k_x^2+\frac{2n}{l_0^2}}}.
\end{eqnarray}

Matrix elements
\begin{eqnarray}\label{f11}
\langle\psi_{m,d'}^{\lambda'}|\hat{v}_z|\psi_{n,d}^{\lambda}\rangle=iv\times\nonumber\\
\Big[\left(a_{3,m,d'}^{\lambda'*}a_{1,n,d}^{\lambda}+a_{4,m,d'}^{\lambda'*}a_{2,n,d}^{\lambda}\right)\delta_{m,n-1}-\\
\left(a_{1,m,d'}^{\lambda'*}a_{3,n,d}^{\lambda}+a_{2,m,d'}^{\lambda'*}a_{4,n,d}^{\lambda}\right)\delta_{m-1,n}\Big].\nonumber
\end{eqnarray}
\begin{widetext}
\begin{eqnarray}\label{f12}
\Re[\sigma_{zz}(\omega)]=\frac{e^2v^2}{2\omega}\Bigg\{
\frac{[1+\cos\alpha_1(k_0)]^2}{|\nu_0'(k_0)|}\Theta\bigg(\omega-\frac{\sqrt{2}v}{l_0}\bigg)\left[f(E_0^{-}(k_0))-f(E_1^{+}(k_0))\right]+\nonumber\\
\frac{[1-\cos\alpha_1(k_0)]^2}{2|\eta_0'(k_0)|}\Theta\bigg(\frac{\sqrt{2}v}{l_0}-\omega\bigg)\left[f(E_0^{-}(k_0))-f(E_1^{+}(k_0))\right]+\nonumber\\
\sum\limits_{n\geq 1}\frac{[1-\cos^2\alpha_n(k_n)][1+\cos^2\alpha_{n+1}(k_n)]}{|\nu_n'(k_n)|}\Theta\bigg(\omega-\frac{\sqrt{2}v}{l_0}\Big(\sqrt{n}+\sqrt{n+1}\Big)\bigg)\left[f(E_n^{-}(k_n))-f(E_{n+1}^{+}(k_n))\right]+\nonumber\\
\frac{[1-\cos^2\alpha_n(k_n)][1+\cos^2\alpha_{n+1}(k_n)]}{2|\eta_n'(k_n)|}\Theta\bigg(\frac{\sqrt{2}v}{l_0}-\omega\bigg)\left[f(E_n^{+}(k_n))-f(E_{n+1}^{+}(k_n))\right]
\Bigg\}.\nonumber
\end{eqnarray}
\end{widetext}
Here $\nu_n(k)=\omega-v\Big(\sqrt{k^2+\frac{2(n+1)}{l_0^2}}+\sqrt{k^2+\frac{2n}{l_0^2}}\Big)$, $\eta_n(k)=\omega-v\Big(\sqrt{k^2+\frac{2(n+1)}{l_0^2}}-\sqrt{k^2+\frac{2n}{l_0^2}}\Big)$, $k_n=\sqrt{\big(\frac{\omega}{2v}\big)^2-\frac{2n+1}{l_0^2}+\big(\frac{v}{\omega l_0^2}\big)^2}$.


\begin{thebibliography}{99}

\bibitem{hasan} M.Z. Hasan and C.L. Kane, Rev. Mod. Phys. {\bf 82}, 3045 (2010).

\bibitem{fu2} L. Fu and C.L. Kane, Phys. Rev. B {\bf 76}, 045302 (2007).

\bibitem{konig} M. K\"{o}nig, S. Wiedmann, C. Br\"{u}ne, A. Roth, H. Buhmann, L.W. Molenkamp, X.L. Qi, and S.C. Zhang, Science {\bf 318}, 766 (2007).

\bibitem{qi} X.-L. Qi and S.-C. Zhang, Rev. Mod. Phys. {\bf 83}, 1057 (2011).


\bibitem{bhz} B.A. Bernevig, T.L. Hughes, and S.-C. Zhang, Science {\bf 314}, no. 5806, 1757 (2006).

\bibitem{vp} B.A. Volkov and O.A. Pankratov, Pis’ma Zh. Eksp. Teor. Fiz. {\bf 42}, 145 (1985) [JETP Lett. {\bf 42}, 178 (1985)].

\bibitem{molenkamp} A. Roth, C. Br\"{u}ne, H. Buhmann, L.W. Molenkamp, J. Maciejko, X.-L. Qi, and S.-C. Zhang, Science {\bf 325}, 294 (2009).

\bibitem{kvon} G.M. Gusev, Z.D. Kvon, O.A. Shegai, N.N. Mikhailov, S.A. Dvoretsky, and J.C. Portal, Phys. Rev. B {\bf 84}, 121302(R) (2011).

\bibitem{buhmann} H. Buhmann, J. Appl. Phys. {\bf 109}, 102409 (2011).


\bibitem{ent-mag} M.V. Entin, and L.I. Magarill, Pis’ma v ZhETF {\bf 103}, 804 (2016) [JETP Lett. {\bf 103}, 711 (2016)].

\bibitem{mah-mag-ent} M.M. Mahmoodian, L.I. Magarill, and M.V. Entin, J. Phys.: Condens. Matter {\bf 29}, 435303 (2017).

\bibitem{mah-ent} M.M. Mahmoodian and M.V. Entin, Phys. Status Solidi B {\bf 256}, 1800652 (2019).

\bibitem{kvon1} A. Rahim, A.D. Levin, G.M. Gusev, Z.D. Kvon, E.B. Olshanetsky, N.N. Mikhailov, and S.A. Dvoretsky, 2D Mater. {\bf 2}, 044015 (2015).

\bibitem{ent-brag1} M.V. Entin and L. Braginsky, Europhys. Lett. {\bf 120}, 17003 (2017).


\bibitem{tchou} S. Tchoumakov, V. Jouffrey, A. Inhofer, E. Bocquillon, B. Placais, D. Carpentier, and M.O. Goerbig, Phys. Rev. B {\bf 96}, 201302(R) (2017).

\bibitem{lu} X. Lu and M.O. Goerbig, Europhys. Lett. {\bf 126}, 67004 (2019).

\bibitem{mah-ent1} M.M. Mahmoodian and M.V. Entin, Phys. Rev. B {\bf 101}, 125415 (2020).

\bibitem{fzhang} F. Zhang, C.L. Kane and E.J. Mele, Phys. Rev. B {\bf 86} 081303 (2012).

\bibitem{shen} S.-Q. Shen, Topological insulators, Springer Series in Solid-State Sciences Vol. 187 (Springer, New York, 2017).

\end{thebibliography}
\end{document}